\documentstyle[12pt]{article}
\begin{document}
\begin{center}
{\Large Achieving pole-law inflation: the extreme inflation }\\
\bigskip
{\bf D.H. Coule}\\
\bigskip
School of Mathematical Sciences\\
University of Portsmouth, Mercantile House\\ 
Hampshire Terrace, Portsmouth PO1 2EG
\bigskip
\begin{abstract}
The pre-big bang's 
inflationary mechanism, when allowance is made for the rapid
change of Newton's constant, 
is not actually of pole-law form . We give examples where
pole-law inflation, which requires violation of the weak-energy
condition, is possible but unlikely due to its very unstable 
character. 

\end{abstract}

PACS numbers: 04.20. Ex, 02.30.

\end{center}
\newpage
{\bf  Introduction}

The pre-big bang scenario, inspired by superstring theory, 
is claimed to be an alternative inflationary universe
model to that of the usual scalar potential 
driven one - for reviews see [1]. The expansion is said to
now start at time $t\rightarrow -\infty$ and is 
of the form $a\sim (-t) ^p$ with $p<0$, often 
dubbed ``super-inflation'' in a
string theory context where $p=-1/\sqrt{3}$ [1]. Later a branch change is
expected to switch to the post-big bang, now a  
less expansive $a\sim t^{1/\sqrt{3}}$ solution,
which could easily join to a conventional FRW expansion. But 
this is also  just an example of 
kinetic or pole-law inflation which is expected to be a more  
general phenomena that occurs in various alternative gravity theories,
and was earlier obtained with an 
induced gravity model [2], which in turn can be
related to compactification of higher dimensional gravity theories [3]. 

Actually 
these claims of pole-law inflation occuring in alternative 
gravity models, and similarly in the pre-big bang case, are erroneous 
as I have explained in a previous paper [4]. 
There I suggested the pre-big bang solution was not 
actually inflationary, but
rather was a contraction with respect to the Planck length: all the  
usual pole-law schemes had ignored this problem about the growing Planck 
length.  Here I wish
to explain what actually would need to  be done to obtain  pole-law
inflationary behaviour and give some possible matter sources: some of which 
have already cropped up in certain string theories. 
By this means I hope to strengthen and clarify
my criticism of the pre-big bang scenario which, with dilaton alone, is 
not an example of this pole-law type. 

{\bf Failure of dilaton driven pre-big bang inflation}

Consider a flat FRW universe model with a perfect fluid equation of state
$p=(\gamma-1)\rho$, the expansion is $a\sim |t|^{2/3\gamma}$ . Now notice
the string theory value $a\sim |t| ^{-1/\sqrt{3}}$ is equivalent to a value
of $\gamma=-2\sqrt{3}/3 \simeq -1$. Unlike conventional inflation which
violates the strong-energy condition $0\leq \gamma<2/3$, we now have the
stricter requirement of violating the weak-energy condition $\gamma<0$, see
eg.[5]. 
 So the string theory violates the weak-energy condition and so it
 explains why pole-law inflation is present ? Well no, we should be 
 careful since in string 
 theory the dilaton causes Newton's constant $G$ to run while the
 energy conditions are formulated in general relativity where the
 Newton's constant is really constant. But we can transform the   
 string theory effective action to one where Newton's constant is fixed
 in the so-called Einstein frame. 

Consider the
simple model that comes from the low-energy string theory
with action [6]
\begin{equation} 
S=\int d^4x \sqrt{-g}\; \exp(-\phi) \left ( R-\omega  (\partial_{\mu}\phi)^2  
\right )\;\;.
\end{equation}
  We have only included the 
  dilaton $\phi$ term as this is the fundamental component that 
  is suggested can possibly
  drive an expansion. Including a cut off for the finite 
  string size, as done by ref.[7], would not be 
  relevant as to whether inflation occurs to  
  give a large universe.\footnote{ One can alternatively formulate the 
 question as to whether inflation can 
 drive the scale factor  
  massively bigger than 
  the string length scale $l_s$ given the value 
  of a fixed  Newton's constant (here $l_s$  is related to the 
  Planck length by the string coupling 
  constant $g$ i.e. $l_p=g l_s$ but which 
  are equivalent  for strong coupling $g\sim 0(1)$ when inflation 
  should finish and a branch change occur).} But of course ultimately one 
  should work with the full string theory when properly formulated in 10, 11
  dimensions. 
  We note in passing that the dilaton's role appears weakened 
  in 11 dimensinal supergravity which might 
  allow a more conventional inflation cf.[8]. An inflation driving 
  field can now remain uncoupled to the dilaton and so keep its `shallowness',
  so allowing  a violation of the strong-energy condition. 

  In the mean time we stick with investigating the dilaton,  
  which is thought a fairly general aspect of strings when starting from 
  10 dimensions. 

  Using a field redefinition $\Phi=\exp(-\phi)$ the
  action can be rewritten in the more usual Brans-Dicke form
\begin{equation}
S=\int d^4x\sqrt{-g}\left ( \Phi R-\frac{\omega}{\Phi}(\partial _
{\mu}\Phi)^2\right )\;\;.
\end{equation}
  Duality symmetry of string theory requires
  $\omega=-1$ [9] but we keep this $\omega$ term general for now. 

   By means of a conformal transformation to new quantities denoted by 
   tildes, such that the new metric becomes-see eg. [10,11], 
    \begin{equation}
\tilde{g}_{\mu\nu}=\Omega^2 g_{\mu\nu}
\end{equation}
and where $\Omega^2=\Phi$ ,
we can find an equivalent action to expression (2). This 
can be expressed as
\begin{equation}
S=\int d^4x \sqrt{-\tilde{g}} \left ( R(\tilde{g}) -1/2 (\tilde{\nabla}
\sigma)^2\right )\;\;,
\end{equation}
where the scalar field $\sigma$ is  defined from [10,11]
\begin{equation}
\Phi=\exp (\beta\sigma)\;\;,
\end{equation}
 and  $\beta^2=1/(2\omega+3)$.
 This action (5) is simply that of a massless scalar field 
 whose field equations with a FRW metric are,
 \begin{equation}
 H^2 = \dot{\sigma}^2
 \end{equation}
 \begin{equation}
 \ddot{\sigma} +3\frac{\dot{a}}{a} \dot{\sigma} =0\;\;\; \Rightarrow \;\;
 \dot{\sigma}^2 =\frac{A}{a^6} \;\;\;(\rm{A=constant})
 \end{equation}

  Newton's constant is now fixed (to unity) and the matter field is just 
   that of a stiff equation of state $\gamma=2$. So we have now apparently 
   lost the presence of
   $\gamma<0$ that is expected to  drive a pole-law inflation. 
   So in the original string frame we have $\gamma\simeq -1$, while 
   in the Einstein frame $\gamma=2$, so we have inflation in the string frame only 
   and thats enough surely? Well not really, recall that inflation requires
   gravity to become repulsive unlike its usually attractive behaviour.
   This I contend should be a fundamental requisite when claiming inflation 
   is present. Further, this aspect of 
   inflation should be conformally invariant.
   Now why do I say gravity is remaining attractive? Newton's constant
   $G$ in the pre-big bang phase starts small and grows as the dilaton
   increases [1,4] such that 
   $G=\exp(\phi)$ so always being positive and since
   there is no cosmological constant (or dilaton potential) 
   present it has no possibility of
   being overwhelmed by any repulsive component. As gravity is attractive
   there is no possibility of naturally getting a large universe and
   by comparing the Planck length scale with the scale factor one
   finds indeed that the universe is collapsing with respect to this 
   Planck length, or alternatively a failure to grow  
   w.r.t. the string length scale [4]. Any ensuing branch change 
   to the post-big bang phase would still require the fixing of arbitrary 
   constants (like $A$ above) 
   to be large to `force' the well known mismatch of scales  
   in the usual big bang model: or else requiring a 
   further period of inflation, so really
   negating the reason for the pre-big bang phase.

   {\bf Use of collapsing universe phase}
   
   Althought collapsing universes do not inflate they do
   provide the possibility of providing a suitable fluctuation spectrum
   for later galaxy formation in the following expansionary phase. 
   By altering
   the matter content one can change the density fluctuation spectrum. As
   we have seen the pre-big bang is a stiff fluid which gives a ``blue
   spectrum'': larger fluctuations on smaller scales [12]. By reducing 
   $\gamma$ one obtains more power on larger
   scales and by the time of a dust ($\gamma=1$) matter source 
   one is getting a more favoured
   scale invariant spectrum [13]. However the use of contraction 
   phases just transforms the 
   `fine tuning' problems of the usual big bang to an equivalent 
   earlier question
   of why the universe started large and with the further requirement that
   one needs a bounce to connect to an expansion phase cf.[4]. Likewise,
   contraction also explains the resolution 
   of the horizon and flatness problems, but
   only by fiat, by starting initially with a 
   large homogeneous universe cf.[14]. 
   
   In summary, the dilaton driven phase is not inflationary in the
   sense of generating a large universe. Rather it is just 
   contracting w.r.t. Planck or string scale, which still begs the 
   question why did the universe start so large
   initially?  Collapsing solutions do allow fluctuations to `leave the
   horizon' and seed future galaxy formation in an ensuing expansionary
   phase.  By altering the equation of state in the 
   collapsing phase, now for matter satisfying 
   the strong-energy condition, the spectrum can be altered to fit
   requirements.
    
{\bf True pole-law inflation}

    How do we obtain a real pole-law inflation and not just
    a contraction phase. Recall that for a 
    scalar field $\phi$ the definition of $\gamma$ is given by
    \begin{equation}
    \gamma =\frac{2\dot{\phi}^2}{\dot{\phi}^2+V(\phi)}
    \end{equation}
    so to obtain $\gamma<0$ we require, 
    since $\dot{\phi}^2>0$ that $V(\phi)<0$
    , that is a negative potential with $|V(\phi)|>\dot{\phi}^2$. By using
    the ideas of ref.[15,16] that one can run Einstein's equations in reverse
    by first fixing the required behaviour of the scale factor, here say
    $a\sim |t|^{-1}$, we can derive the required scalar potential. Here it
    would be  of the form of a negative expontial $V(\phi) \sim -\exp(\phi)$. 
    Such a simulation only applies for a certain time before the field
    falls away to $\phi \rightarrow \infty$ [16,17]. 
    This behaviour can be contrasted
    with an open anti-DeSitter space where the scale factor 
    goes like $a\sim cos(t)$, where there is also some expansion for time 
    going from $-\pi/2$ to zero, and indeed anti-DeSitter also violates the
    weak-energy condition. 

    Negative exponential potentials might occur in the low energy limit
    of 11 dimensional supergravity theories cf. [18]

    Some other sources can violate the weak-energy condition, for example
    the Brans-Dicke model but now with $\omega<-3/2$. 
    The original induced gravity matter source [2]  that was supposed to give
    pole-law inflation also only 
    caused a contraction \footnote{ It is straightforward to show, using the
    same argument as ref. [4],
    that this contraction occurs for any 
    $p<1$ so including all the pole-law behaviour $p<0$.} since only values
    of the parameter $\epsilon>0$ that gave conformal 
    equivalence to a massless
    scalar field chosen. Taking more extreme values of $\epsilon$ or
    $\omega<-3/2$ seems to allow weak-energy condition violation 
    but are unlikely to give stable solutions [11]. 
    These weak-energy violating values 
    have also been considered for the support of transversable
    Lorentzian wormholes [19] 
    which by careful engineering can be made stable, 
    but this is hardly feasible for  cosmological solutions.

    Higher order gravity theories can give  
    negative potential in their conformally related scalar field models
    provided one takes `wrong signs' in front of the coefficients of the
    higher order Ricci scalar eg. $R^2$ terms [20,21]. 
    
    Also 
    Lagrangians of the form ${\cal L} =\ln (1+R)$ or
    $\exp(\lambda R)$, give negative, often exponential, potentials [20]. 
    Certain non-analytic functions of Ricci tensors terms might be also be 
    applicable [22]. 
    
    Higher order corrections to string theory give Gauss-Bonnet terms that
    were hoped could amend the end of any pre-big bang phase [23]. 
    These terms can also violate 
    the weak-energy condition but are known to be unstable [11,24],
   especially  because of Ricci tensor squared terms 
    which are susceptible to growing anisotropy divergences [25]: 
    they further apparently  cause runaway black hole production [23]. 
    
    It would seem
    preferable to just work with higher order corrections  
    with `correct signs'
    that can still violate the strong-energy condition: 
    such schemes have been
    presented in [26].

     A non-minimally coupled scalar field with sufficiently large field
     causes a violation of the weak-energy condition in that Newton's
     constant effectively changes sign. The so-called rebouncing behaviour
     of ref.[27] is closely related to the possibility of pole-law expansion.

    Bulk viscosity could in theory 
    gives an effective $\gamma<0$, see eg.[28], 
    althought it is suggested that it might 
    not even give a negative pressure i.e. $\gamma<2/3$ [29]. 
    
    Coupling 
    together of scalar fields such as a dilaton and axion field [30] can
    also violate the weak-energy condition, there they obtained an
    effective equation of state $\gamma=-2$ and worked only with the rapid
    contraction phase which occurs for positive $t>0$,  so still 
    starting with a usual, time equal to zero, big bang model.

     Trying to work with weak-energy violating inflation is likely to be
   fraught with instabilities cf.[11] and possible 
   divergences in anisotropies
    as in the non-minimally coupled scalar field [27].  
    To use a particular example one would need to consider the stability   
    carefully  and arrange a suitable mechanism to end such inflation: at
    present no example seems remotely realistic.
    
    In Fig.(1) we follow
    how the expansion rate becomes faster as the strong-energy condition
  is violated in conventional inflationary models. 
  As the weak-energy condition
    also becomes violated the behaviour `switches over': expansion now 
    at negative times
     and rapid contraction for positive times. Incidentally, 
     this is an example of 
     $\em fragility$ when going to negative $\gamma$: small changes 
     in equation of state giving large changes in behaviour [31,16]. The
     extreme expansion in the 
     scale factor only happens for $|t|<1$ so requiring going 
     infinitesimally close to the still singular behaviour (  $ \dot{a} 
     \rightarrow \infty $ as $ t \rightarrow 0_{-} $ ) at
     time zero, here an arbitrary constant. One is needing to work 
     within the quantum gravitational regime to drive the expansion: this
     requires a correct version of quantum gravity, here string theory, 
     unlike ordinary inflation
     which can still be relied on to proceed for below Planck scale values.

      If one wishes to use the expansion
     at negative times you still need 
     to match to a regular expansion now at
     positive time. Getting from one 
     domain to another seems an added complication
    as the time parameter has to be 
    carefully matched during this transition.
     If the weak-energy condition becomes 
     satisfied too soon the universe will
    start collapsing for negative times 
    (or in general for time below an arbitrary 
     constant). What sets the time parameter 
     correctly is an added complication
     over regular potential driven inflation, although quantum gravitational
	  effects might give some hope of its resolution. The growing 
	  Hubble parameter, also a feature of weak-energy violating
	  inflation, is also a problem since there are strong limits on
	  the allowed size of the Hubble parameter to avoid gravitational
	  waves [32] and the growing Hubble parameter will still  
	  eventually result in a curvature singularity. 
	  
	  Although 
     weak-energy violating inflation remains a 
     method of last resort, with its 
     extreme expansion and tendency for instabilities, we suspect it 
      is likely to prove `too hot to handle'. Searching string theory
      for strong-energy violating inflationary mechanisms, either by
      preventing the dilaton from rolling, or from higher order gravity
      corrections with stable signs, seems a more realistic endeavor.

{\bf Acknowledgement}\\
I should like to thank participants of the UK HEP institute (Oxford 98) for
various helpful discussions about string cosmologies.

\newpage
{\large Figure Captions}\\
{\bf Figure 1)}.\\ 
{\em Expansion rate for various matter sources}  \\
The scale factor $a$ against time  $t$ is plotted for matter sources
ranging from strong-energy condition being satisfied (I) to weak-energy
condition being violated (IV). Plot (I) radiation 
$(\gamma=4/3)$ : (II) coasting solution $(\gamma=2/3)$, where strong-energy
condition is just being violated : (III) exponential expansion $(\gamma=0)$,
the most extreme inflation with only strong-energy condition violated :
(IV) Pole-law expansion $(\gamma=-2/3)$, now weak-energy condition is 
violated to give extreme expansion for negative times or rapid contraction
for positive times. Note that use of pole-law expansion to generate 
a large scale factor requires going closer  than Planck time to the 
singular behaviour at, here, $t=0$. Before a switch to  
a more normal expansionary behaviour can intervene.

\newpage

{\bf References}\\
\begin{enumerate}
\item M. Gasperini and G. Veneziano, Astropart. Phys. 1 (1993) p.317.\\
M. Gasperini,``Birth of the Universe in String cosmology'', preprint 
gr-qc/9706037.\\
S.J. Rey, ``Recent progress in string inflationary cosmology'',
preprint hep-th/9609115\\
J.J. Levin, ``Gravity driven inflation'', preprint gr-qc/9506017.\\
Available at pre-big bang web site:\\ 
http://www.to.infn.it/teorici/gasperini/
\item M.D. Pollock and D. Shahdev, Phys. Lett. B 222 (1989) p.12.
\item D. Sahdev, Phys. Lett. B 137 (1984) p.155.\\
Phys. Rev. D 30 (1984) p.2495.
\item D.H. Coule, Class. Quant. Grav. 15 (1998) p.2803. 
\item S.W. Hawking and G.F.R. Ellis, ``The large scale structure
of Space-time'',( Cambridge: Cambridge University Press) 1973.\\
M. Visser, ``Lorentzian Wormholes'', (New York: AIP Press) 1995.
\item E.S. Fradkin and A.A. Tseytlin, Phys. Lett. B 158 (1985) p.316.\\
C.G. Callan, D. Friedan, E.J. Martinec and M.J. Perry, Nucl. Phys. B 262
(1985) p.593.\\
C. Lovelace, Nucl. Phys. B 273 (1985) p.135.
\item  N. Kaloper, A. Linde and R. Bousso, preprint hep-th/9801073. 
\item N. Kaloper, I.I. Kogan and K.O. Olive,  Phys. Rev. D 57 (1998) p.6242.
\item G. Veneziano, Phys. Lett. B 263 (1991) p. 287.
\item S. Kalara, N. Kaloper and K.A. Olive, Nucl. Phys. B 341 (1990) p.252.
\item G. Magnano and L.M. Sokolowski, Phys. Rev. D 50 (1994) p. 5039.\\
L.M. Sokolowski, preprint gr-qc/9511073.
\item J. Hwang and H. Noh, Phys. Rev. D 54 (1996) p. 1460\\
\item M. Gasperini and G. Veneziano, Mod. Phys. Lett. A 8 (1993) p.3701.
\item D. Wands, preprint gr-qc/9809062.
\item G.F.R. Ellis and M.D. Madsen, Class. Quant. Grav. 8 (1991) p.667.
\item G.F.R. Ellis, J.E.F. Skea and R.K. Tavakol, Europhys. Lett. 16 (1991)
p.767.
\item G.F.R. Ellis, D.H. Lyth and M. Miji\'{c}, Phys. Lett. B 271 (1991)
p.52.
\item M.S. Bremer, M.J. Duff, H. Lu, C.N. Pope and K.S. Stelle,
preprint hep-th/9807051\\
S.W. Hawking and H.S. Reall, preprint hep-th/9807100.
\item K.K. Nandi and A. Islam, Phys. Rev. D 55 (1997) p.2497.\\
A.G. Agnese and M. La. Camera, Phys. Rev. D 51 (1995) p.2011.
\item J.D. Barrow and S. Cotsakis, Phys. Lett. B 214 (1988) p.515.\\
{\em ibid}, Phys. Lett. B 258 (1991) p.299.
\item K. Maeda, Phys. Rev. D 37 (1989) p.858.
\item E. Br\"{u}ning, D.H. Coule and C. Xu, Gen. Rel. Grav. 26 (1994) 
p.1193.
\item S. Kawai, M. Sakagami and J. Soda, preprint gr-qc/9802033.
\item L.M. Sokolowski, Z. Golda, M. Litterio and L. Amendola, Int. J.
Mod. Phys. A 6 (1991) p.4517.
\item A.L. Berkin, Phys. Rev. D 44 (1991) p.1020.
\item J. Ellis, N. Kaloper, K.A. Olive and J. Yokoyama, preprint
hep-th/9807482.
\item T. Morishima and T. Futamase, preprint gr-qc/9808040.
\item R. Maartens, Class. Quant. Grav. 12 (1995) p.1455.\\
W. Zimdahl, Phys. Rev. D 53 (1996) p.5483.
\item T. Pacher, J.A. Stein-Schabes, M.S. Turner, Phys. Rev. D 36
(1987) p.1603.
\item F.G. Alvarenga and J.C. Fabris, Class. Quant. Grav. 12 (1995)
p. L69.
\item R.K. Tavakol and G.F.R. Ellis, Phys. Lett. A 130 (1987) p.217.
\item  V.A. Rubakov, M.V. Sazhin and A.V. Veryaskin, 
Phys Lett. B 115 (1982) p.189.\\
L.F. Abbott and M.B. Wise, Nucl. Phys. B 244 (1984) p.541.
\end{enumerate}
\end{document}